\documentclass[runningheads]{llncs}
\usepackage[T1]{fontenc}
\usepackage{graphicx}
\usepackage{multirow}
\usepackage{balance} 
\usepackage{stfloats}

\usepackage{amssymb,amsfonts}
\usepackage[linesnumbered,ruled,vlined]{algorithm2e}
\SetAlFnt{\footnotesize}

\SetCommentSty{mycommfont}
\usepackage{longtable}
\usepackage{pdflscape}
\usepackage{tcolorbox}
\SetKwInput{KwInput}{Input}                
\SetKwInput{KwOutput}{Output}
\usepackage{textcomp}
\usepackage{xcolor}
\usepackage{hyperref}

\usepackage{multirow}
\usepackage{float}
\usepackage{booktabs}
\usepackage{dcolumn}
\usepackage{rotating} 
\usepackage{arydshln}
\usepackage{listings}
\usepackage{amssymb}
\usepackage{graphicx}
\usepackage{wrapfig}
\usepackage{balance}
\usepackage{url}
\usepackage{hhline}
\usepackage{xcolor}
\usepackage{multirow}
\urldef{\mailsa}\path||    
\usepackage{color}
\usepackage{tikz}

\usepackage{adjustbox}
\usetikzlibrary{calc, shapes.geometric, arrows, positioning}
\definecolor{mygreen}{rgb}{0,0.6,0}
\definecolor{mygray}{rgb}{0.5,0.5,0.5}
\definecolor{mymauve}{rgb}{0.58,0,0.82}

\begin{document}



\title{A Systematic Study of Deprecated Android API Usage Updates}

\author{Tarek Mahmud\inst{1} \and
Meiru Che\inst{2} \and
Guowei Yang\inst{3}}
\authorrunning{T. Mahmud et al.}
%
\institute{
Texas A\&M University--Kingsville, Kingsville, TX, USA\\
\email{Tarek.Mahmud@tamuk.edu}
\and
Central Queensland University, Brisbane, QLD, Australia\\
\email{m.che@cqu.edu.au}
\and
The University of Queensland, Brisbane, QLD, Australia\\
\email{guowei.yang@uq.edu.au}
}

\maketitle

\begin{abstract}
Android Application Programming Interfaces (APIs) continuously evolve, causing existing APIs to become deprecated and requiring developers to migrate their usages to recommended replacements. Although prior studies have proposed automated migration techniques, their effectiveness is often limited by the diversity and contextual complexity of API changes. This paper presents a systematic study of deprecated Android APIs, their recommended replacements, and real-world usage updates performed by developers. Based on an analysis of 1,000 deprecated APIs and 4,336 migration examples collected from GitHub, we identify six migration categories and derive recurring usage-update patterns involving Android-version checks, parameter adaptation, object initialization, data-type conversion, and changes in API interaction mechanisms.

To assess their applicability, we augment the zero-shot GUPPY benchmark \cite{mahmud2025automated} prompts with the migration category and corresponding usage-update pattern. On 156 benchmark tasks, the pattern-guided prompt improves correct updates from 139 to 147 and performs better in five of the six categories.
\end{abstract}




\section{Introduction} 
\label{introduction}
In recent years, mobile devices have become indispensable in daily life, supporting activities such as navigation, communication, education, and entertainment. Among mobile operating systems, Android dominates the market with over 70\% share~\cite{market-share} and more than 3.2 billion active devices as of 2022~\cite{goog}. As of 2023, the Google Play Store hosted over 2.69 million Android apps~\cite{app}, built using the Android Software Development Kit (SDK) and its extensive set of APIs. However, as Android evolves, its APIs are frequently updated by introducing new methods, modifying existing interfaces, and deprecating outdated functionality. The current Android API level has reached 34, underscoring the continuous evolution of the platform.

Android API deprecation marks APIs as outdated and signals that developers should migrate to recommended replacement APIs when available. Although deprecated APIs may remain functional on older Android versions, their continued use can lead to compatibility issues, crashes, feature loss, security concerns, or maintenance difficulties as the platform evolves. To manage these changes, the Android framework follows a structured deprecate-replace-remove cycle~\cite{li2018characterising}. In this process, an API is first marked as deprecated in the SDK documentation, for example through the \texttt{@Deprecated} annotation, and is often accompanied by a deprecation message that recommends an alternative. For instance, the deprecation message for \texttt{getAllNetworkInfo()} recommends using \texttt{getAllNetworks()} instead. After a grace period that may span several Android releases, deprecated APIs can be removed from the framework. This phased process gives developers time to update their applications while preserving compatibility across Android versions. Consequently, updating deprecated API usages is important for maintaining app compatibility, adopting newer platform functionality, and preserving application quality.

Prior research has investigated Android API evolution~\cite{mcdonnell2013empirical,linares2013api,linares2014api,li2018characterising,apicia,lamothe2021systematic} and proposed automated techniques for updating deprecated API usages~\cite{fazzini2019automated,haryono2020automatic,haryono2021androevolve,zhao2022towards,mahmud2025automated}. These techniques have demonstrated the potential of example-based transformation, program analysis, handcrafted templates, and large language models for automating API migration. However, existing approaches mainly focus on generating updates rather than systematically understanding how developers perform such migrations in practice. In particular, limited attention has been paid to the recurring structural and behavioral patterns that developers follow when adapting parameters, handling Android-version differences, creating replacement objects, converting data types, or restructuring API interactions. This lack of systematic migration knowledge can limit the effectiveness and generalizability of automated approaches, especially when replacement APIs require more than a direct syntactic substitution.

In this paper, we systematically analyze Android's officially documented deprecated APIs, their deprecation messages, and their recommended replacement APIs. We further examine real-world code changes to understand how developers update deprecated API usages in Android applications. From this analysis, we identify six migration categories and derive reusable usage-update patterns that capture recurring developer strategies, including Android-version checks, parameter adaptation, object initialization, data-type conversion, and changes in API interaction mechanisms. Rather than proposing a complete migration tool, our primary goal is to uncover and formalize the strategies developers follow during API migration.

We also assess the applicability of these patterns to automated API updating by augmenting the zero-shot GUPPY ~\cite{mahmud2025automated} prompt with the migration category and corresponding usage-update pattern. On the 156 tasks in the GUPPY benchmark, the pattern-guided prompt increases the number of correct updates from 139 to 147 and performs better in five of the six migration categories. These results show that empirically derived migration patterns can provide useful domain guidance for LLM-based deprecated API updating.

Specifically, we make the following contributions:

\begin{itemize}

\item \textbf{Categorization of Migration Scenarios:} We identify and categorize recurring migration scenarios based on the structural relationships between deprecated and replacement APIs, including changes in return types, parameters, data types, and API kinds.

\item \textbf{API Usage-Update Patterns:} We analyze real-world migration examples and derive common patterns that developers apply when updating deprecated API usages, including version checks, parameter adaptation, object initialization, data-type conversion, and interaction restructuring.

\item \textbf{Applicability study:} We incorporate the identified migration categories and usage-update patterns into the GUPPY prompt and evaluate their usefulness on the GUPPY benchmark~\cite{mahmud2025automated}. The results provide empirical evidence that developer-derived migration patterns can improve LLM-guided deprecated API updates.

\end{itemize}

\section{Background} \label{background}


\subsection{Android API and Android Developers Website} 
The Android operating system, introduced in 2008 by the Open Handset Alliance, is primarily designed for touchscreen mobile devices such as smartphones and tablets. The Android Software Development Kit (SDK) provides libraries, APIs, emulators, debugging tools, and sample projects that allow developers to access platform features including cameras, location services, Wi-Fi, and storage. As Android evolves, new platform versions and API levels are released to introduce functionality, improve performance, and address defects and security concerns.

The Android Developers website, available at \url{https://developer.android.com}, serves as the official documentation portal for Android development. It provides API references, tutorials, design guidance, development tools, and resources for building applications across devices such as phones, tablets, and wearables.

\subsection{Deprecated APIs, Replacement API and API Usage Update}
Android APIs are frequently deprecated as the platform evolves. A deprecated API is an interface whose continued use is discouraged, typically because a newer or more suitable alternative is available. Although deprecated APIs may remain functional for several Android versions, their continued use can introduce compatibility, maintenance, performance, or security concerns. Developers must therefore periodically update their applications to adopt recommended replacement APIs.

Replacement APIs are generally introduced to provide improved functionality, security, efficiency, or compatibility with newer Android versions. Updating a deprecated API usage often requires more than a direct replacement because applications may still support older Android versions where the new API is unavailable. Developers therefore commonly use Android-version checks to invoke the replacement API on newer versions while retaining the deprecated API for older versions.

For example, \texttt{Environment.getExternalStorageDirectory()} was deprecated in favor of \texttt{getExternalFilesDir()}, which provides application-specific access to external storage. A version-compatible update may be structured as follows:

\begin{lstlisting}[caption={How the deprecated API usages should be updated in the app code.}, label=lstExample]
if(Build.VERSION.SDK_INT <= 21) {
	// Deprecated API usage
	File externalStorageDir = 
		Environment.getExternalStorageDirectory();
}
else {
	// Recommended API usage
	File externalFilesDir = 
		getExternalFilesDir(null);
}
\end{lstlisting}

Such updates help applications remain compatible across Android versions while reducing the risks associated with unsupported or outdated APIs.

\newcommand{\rqZero}{RQ1: To what extent do deprecated APIs and their deprecation messages and replacement APIs impact the evolution and usage of Android APIs?}
\newcommand{\rqOne}{RQ1: What are the defining characteristics of the migration process from a deprecated API to its replacement in Android?}
\newcommand{\rqTwo}{RQ2: In real-world apps, how do developers update deprecated API usages?}
\newcommand{\rqThree}{RQ3: To what extent do the identified API usage-update patterns improve LLM-based deprecated API migration?}

\section{Methodology} \label{approach}

This section presents our research questions and describes the data collection,
manual analysis, pattern derivation, and evaluation procedures used in this study.

\subsection{Research Questions}

This study aims to answer the following research questions:

\begin{itemize}

\item \textit{\rqOne} \\ 
By examining the structural relationships between deprecated APIs and their
recommended replacements, we seek to characterize the changes involved in
Android API migration. These relationships include changes in method names,
parameters, return types, data types, classes, packages, and API types.

\item \textit{\rqTwo} \\ 
By analyzing real-world code changes, we investigate how developers update
deprecated API usages for the migration categories identified in RQ1. This
analysis aims to identify recurring strategies, challenges, and usage-update
patterns followed in practice.


\end{itemize}

\subsection{Data Collection}

To answer RQ1, we collect deprecated Android APIs, their deprecation messages,
and their recommended replacement APIs from the official Android Developers
documentation~\cite{androidDevelopers}. We use these deprecated--replacement
API pairs to identify and categorize the structural changes involved in API
migration.

To answer RQ2, we collect real-world API usage-update examples from GitHub for
1,000 randomly selected deprecated APIs. These examples allow us to investigate
how developers implement the migrations identified in RQ1.

\subsubsection{Collection of Deprecated and Replacement APIs}
\label{CollectionOfDeprecatedAPIs}

We begin by extracting Android class names and documentation links from the
Android Developers reference website at
\url{https://developer.android.com/reference/classes}. We retrieve the methods
and fields of each class and identify APIs marked as deprecated. For each
deprecated API, we collect its deprecation message and search the message for a
recommended replacement API.

We identify replacement APIs by matching recurring expressions in the
deprecation messages, including the following:

\begin{itemize}
    \item Use [the] \textlangle a href=``link''\textrangle text\textlangle/a\textrangle
    \item Use [the] \textlangle code\textrangle text\textlangle/code\textrangle
    \item Use [the] text [instead]
    \item Replaced [by][with] text
    \item In favor of text
    \item Rename to text
    \item Call text
    \item Move to [the] text
    \item [may][will] be removed
\end{itemize}

When a pattern is matched, we collect the referenced text and documentation
link and verify that the replacement API exists in the Android Developers
documentation. For API methods, we record the package, class, method name,
return type, and parameter types. For API fields, we record the package, class,
field name, and data type. If a deprecation message recommends multiple APIs,
we collect all of them. If the message indicates that the API will be removed
without recommending an alternative, we classify it as having no direct
replacement.

We evaluated the replacement-extraction procedure using 300 randomly selected
deprecated APIs. Manual inspection identified 231 APIs whose deprecation messages contained replacement APIs. The procedure correctly identified 230 of
them without producing any false positives, resulting in 100\% precision and
99.6\% recall.

\subsection{Identification of API Migration Categories and Usage-Update Patterns}

Two authors independently analyzed the deprecated--replacement API pairs and
real-world code changes. They cross-validated their findings and resolved
disagreements through discussion.

\subsubsection{Manual Analysis of Deprecated and Replacement APIs}

To answer RQ1, the authors examined 1,000 randomly selected deprecated APIs and
their recommended replacements. The sample included at least one API from each
Android class containing deprecated elements.

For each pair, they compared the API type, name, parameters, return type, data
type, class, and package. They identified changes such as renaming, parameter
addition or removal, return- or data-type changes, package relocation, and
transitions between fields and methods.

The relationships were first classified as 1-to-0, 1-to-1, or 1-to-multiple,
and then organized into six migration categories based on the structural
differences between the deprecated and replacement APIs.

\subsubsection{Manual Analysis of Real-World Code Changes}

To answer RQ2, we collected up to five GitHub migration examples for each of
the 1,000 selected APIs, resulting in 4,336 usage-update examples.

The authors analyzed the changes made by developers, including version checks,
API replacement, parameter adaptation, object initialization, data-type
conversion, and changes involving fields, methods, listeners, or callbacks.
Each example was mapped to a migration category, and similar changes were
grouped to derive generalized usage-update patterns.

The patterns were refined iteratively when examples revealed additional
strategies or contextual requirements. The authors then re-evaluated a subset
of examples to verify the consistency of the classifications and patterns.



Two authors independently performed the classification and resolved
disagreements through discussion.

\section{Results and Analysis} \label{evaluation}

\subsection{\rqOne}

To answer RQ1, we analyzed 1,000 randomly selected deprecated APIs and their
recommended replacement APIs. The sample included APIs from different Android
classes and functional domains. For each deprecated--replacement pair, we
examined the number of replacement APIs, API type, method or field name,
parameters, return type, data type, class, and package. This analysis allowed us
to characterize both the structural relationship between the APIs and the
refactorings required during migration.

We first identified three relationships based on the number of deprecated and
replacement APIs involved:

\begin{itemize}
    \item \textbf{1-to-0:} The deprecated API has no recommended replacement.
    In these cases, developers may need to remove the API usage, redesign the
    affected functionality, or continue using the deprecated API when no
    alternative is available.

    \item \textbf{1-to-1:} The deprecated API is replaced by a single API. This
    is the most direct migration scenario, although the replacement may still
    differ in parameters, return type, class, package, or API type.

    \item \textbf{1-to-multiple:} The deprecated API is replaced by two or more
    APIs. These migrations often require multiple calls or additional logic to
    reproduce the behavior of the deprecated API. We did not observe any
    deprecated API with more than three recommended replacements.
\end{itemize}

\begin{figure}[t]
    \centering
    \includegraphics[width=0.6\textwidth]{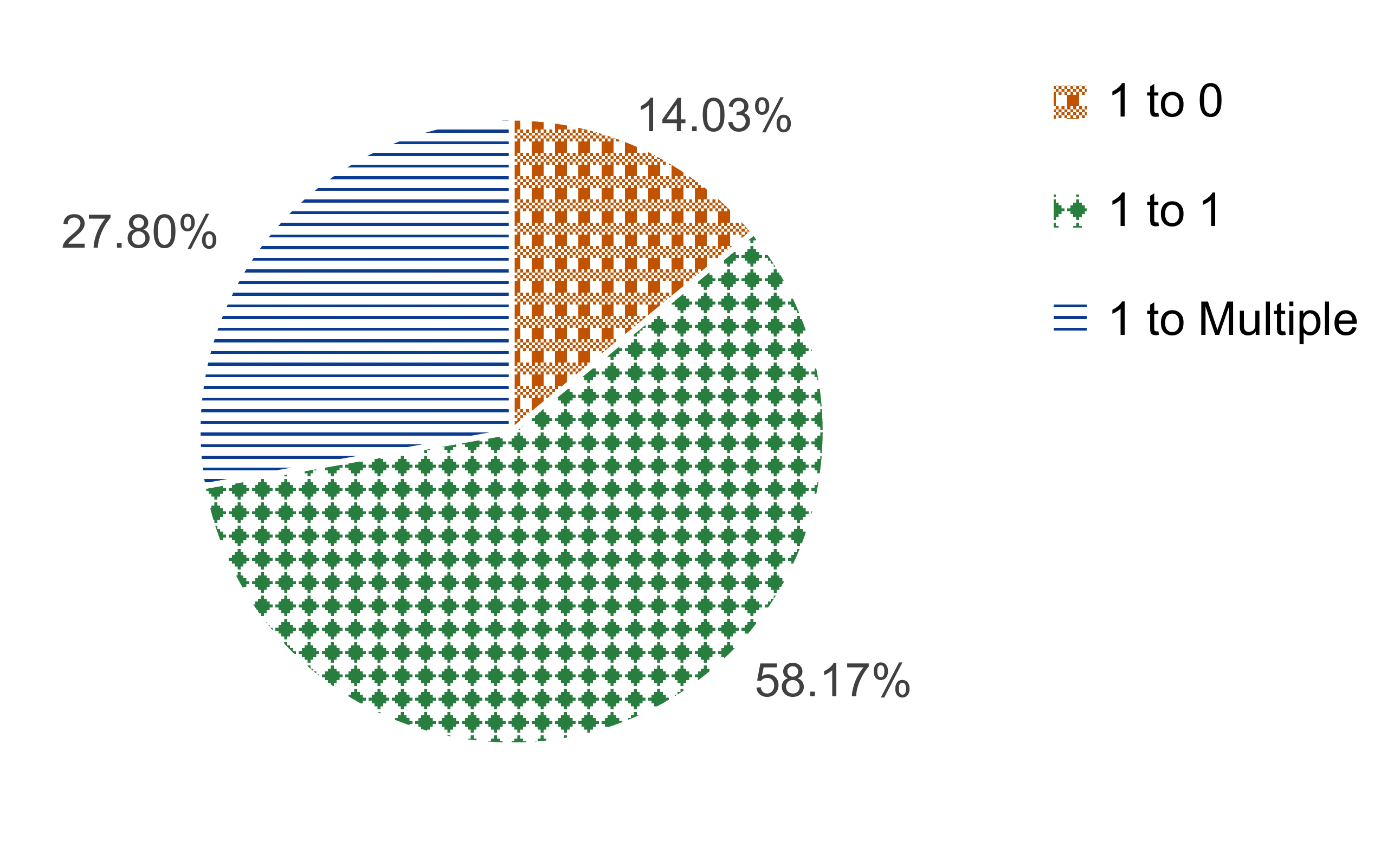}
    \caption{Distribution of API mapping scenarios.}
    \label{fig:api_mapping}
\end{figure}

As shown in Fig.~\ref{fig:api_mapping}, 1-to-1 mappings are the most common,
accounting for 58.17\% of the analyzed APIs. The 1-to-0 and 1-to-multiple
relationships account for 14.03\% and 27.80\%, respectively.

We next compared the signatures of the deprecated and replacement APIs to
identify the refactoring operations involved in the migration. The observed
operations include:

\begin{itemize}
    \item \textbf{Remove API:} Remove the deprecated API when no direct
    replacement is provided.
    \item \textbf{Add or remove parameters:} Modify the parameter list required
    by the replacement method.
    \item \textbf{Rename methods or fields:} Replace an API with another API
    that provides similar functionality under a different name.
    \item \textbf{Change return type or data type:} Adapt the surrounding code
    when the replacement API produces a different type of value.
    \item \textbf{Change class or package:} Access the replacement API through a
    different class, package, or module.
    \item \textbf{Change API type:} Replace a field with a method or a method
    with a field.
\end{itemize}

A migration may involve several of these operations simultaneously. For example,
\url{android.text.format.DateFormat.format(String,long)} was replaced by
\url{java.text.DateFormat.format(Date)}. This migration requires removing and
adding parameters, constructing a \url{Date} object, and changing the class
through which the formatting operation is performed.

Based on these structural differences, we define the following six migration
categories:

\begin{enumerate}
    \item \textbf{Same return type and parameters.}
    The deprecated and replacement methods have the same return type and
    parameters but differ in name, class, or package. For example,
    \url{TimePicker.getCurrentHour()} was replaced by
    \url{TimePicker.getHour()}.

    \item \textbf{Same return type with different parameters.}
    The methods have the same return type but different parameter lists. For
    example, \url{getSystemService(String)} was replaced by
    \url{getSystemService(Class<T>)}.

    \item \textbf{Different return type and parameters.}
    The replacement changes the return type, parameters, or requires multiple
    APIs. For example, \url{ConnectivityManager.getAllNetworkInfo()} was
    replaced by \url{getAllNetworks()} and
    \url{getNetworkInfo(Network)}.

    \item \textbf{API fields of the same data type.}
    A deprecated field is replaced by another field with the same data type.

    \item \textbf{API fields of different data types.}
    A deprecated field is replaced by another field with a different data type.
    For example, \url{Build.VERSION.SDK} was replaced by
    \url{Build.VERSION.SDK_INT}.

    \item \textbf{Different types.}
    A field is replaced by a method or a method is replaced by a field. For
    example, \url{View.SCALE_X} and \url{View.SCALE_Y} were replaced by
    \url{View.setScaleX()} and \url{View.setScaleY()}.
\end{enumerate}

For 1-to-multiple mappings, we assign Category 6 when at least one replacement
has a different API type from the deprecated API. Otherwise, the migration is
assigned to Category 3.

We implemented the category assignment using the deprecated and replacement API
signatures. Regular expressions are used to extract the API type, return type,
data type, and parameter list. For a 1-to-0 mapping, the procedure records that
no replacement exists. For a 1-to-multiple mapping, it examines the types of all
recommended replacements and assigns either Category 3 or Category 6 as
described above.

For 1-to-1 mappings, the procedure first compares the API types. A field-to-method
or method-to-field migration is assigned to Category 6. For method-to-method
migrations, different return types result in Category 3. When the return types
match, the common parameters are removed from both signatures; if no unmatched
parameters remain, the migration is assigned to Category 1, otherwise it is
assigned to Category 2. For field-to-field migrations, matching data types
result in Category 4, while different data types result in Category 5.

\begin{figure}[t]
    \centering
    \includegraphics[width=0.7\textwidth]{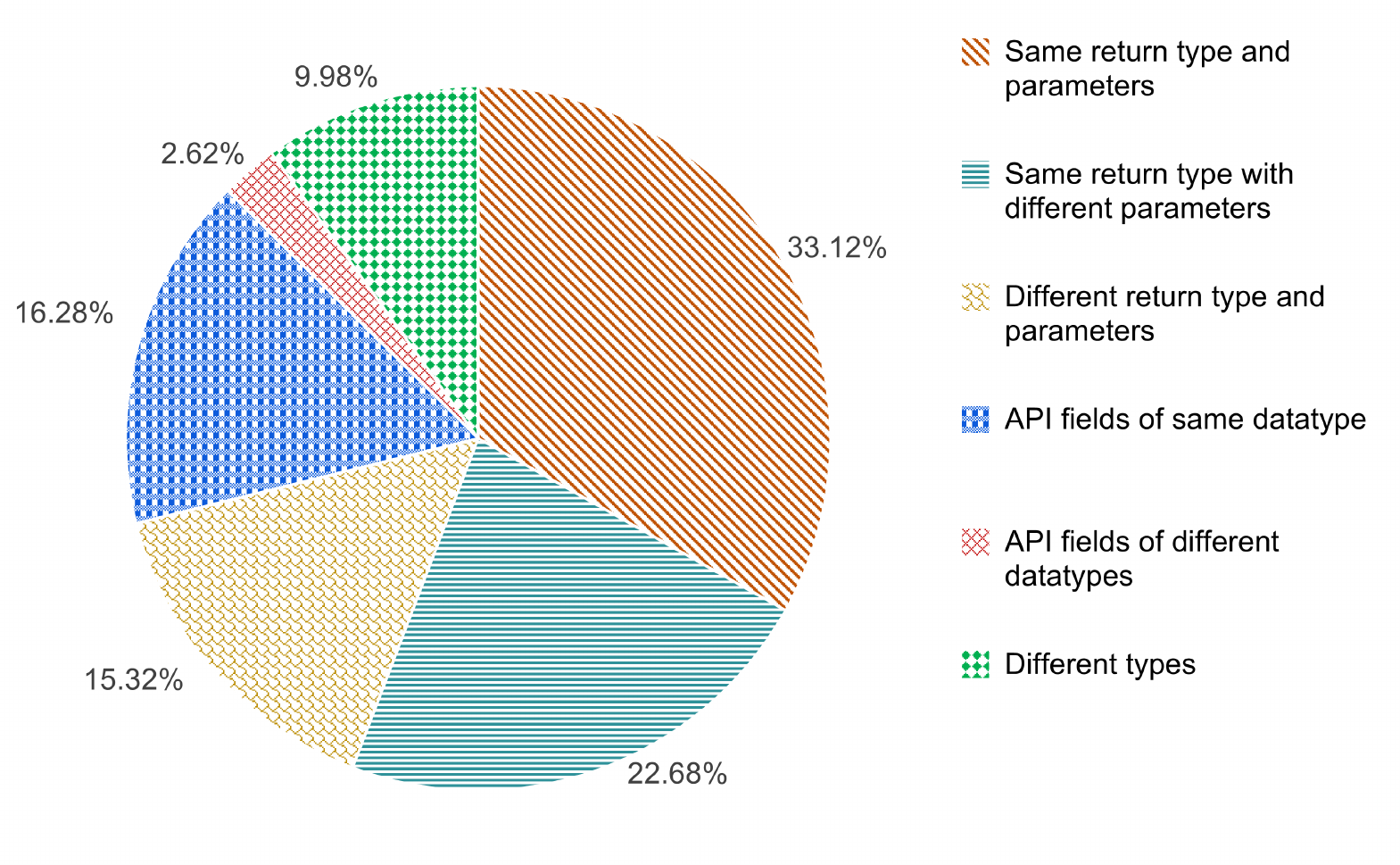}
    \caption{Distribution of the migration categories.}
    \label{fig:categories}
\end{figure}

Fig.~\ref{fig:categories} shows that 33.12\% of migrations have the same return
type and parameters, 22.68\% have the same return type with different
parameters, 15.32\% have different return types or parameters, 16.28\% involve
fields of the same data type, 2.62\% involve fields of different data types,
and 9.98\% involve different API types. Overall, the distribution indicates
that structurally simpler migrations are more common than migrations requiring
substantial type or interaction changes.

\subsection{\rqTwo}
{To answer this research question, we investigated how developers update deprecated API usages in real-world Android apps by analyzing code changes from deprecated to replacement APIs. We utilized the 1,000 randomly selected deprecated APIs from RQ1, collecting up to five update examples per API from GitHub, resulting in a total of 4,336 examples (some APIs had fewer than five updates due to limited availability). These examples were manually analyzed by two of the authors, both experienced in Android development and software evolution research, to ensure consistency and expertise in identifying patterns.

The analysis process involved a systematic, iterative approach. First, we examined each example to identify the specific code changes made (e.g., version checks, parameter adjustments, type conversions). Second, we mapped these changes to the RQ2 categories based on API signature differences (e.g., return type, parameters, field vs. method). Third, we iteratively refined the observed patterns into generalized patterns by grouping similar updates and resolving discrepancies through discussion, achieving a consensus on six mostly used patterns. The criteria for labeling focused on the structural and semantic adjustments developers applied, such as conditional versioning or type handling, ensuring the patterns captured practical migration strategies rather than trivial syntactic fixes.

While these patterns may appear straightforward, their novelty lies in systematically distilling real-world practices into a taxonomy that reveals both commonalities and complexities across diverse API migrations. For instance, the patterns uncover nuanced challenges, like context-dependent decisions in category 3 or semantic shifts in category 6, that go beyond simple replacements, offering insights into developer intent and adaptation strategies not previously formalized.

To assess coverage and representativeness, we quantified the applicability of each pattern across the 4,336 examples:
\begin{itemize}
    \item \textbf{Category 1 (Same return type and parameters):} 1464 examples (33.8\%), reflecting its prevalence due to minimal refactoring needs.
    \item \textbf{Category 2 (Same return type with different parameters):} 1006 examples (23.2\%), common in parameter-evolving APIs.
    \item \textbf{Category 3 (Different return type and parameters):} 656 examples (15.1\%), often involving multiple replacements.
    \item \textbf{Category 4 (API fields of the same data type):} 714 examples (16.5\%), typical for field renames.
    \item \textbf{Category 5 (API fields of different datatypes):} 133 examples (3.1\%), less frequent due to type complexity.
    \item \textbf{Category 6 (Different types):} 363 examples (8.3\%), capturing field-to-method shifts.
\end{itemize}

These distributions align with RQ2 findings (Fig. \ref{fig:categories}), suggesting the patterns are representative of the broader dataset, though rare cases (e.g., category 5) indicate potential gaps in less common scenarios. The patterns cover 100\% of the analyzed examples, as each update fit at least one category, though some complex cases required combining patterns (e.g., category 3 with type conversion from category 5).}

The derived patterns are as follows:

\begin{enumerate}
    \item \textbf{Same return type and parameters.} 
    For this type of change, developers use a straightforward approach, applying version checks to ensure the correct method is invoked for different Android versions. Listing \ref{lstcat1rule} shows the usage update pattern for this category. If the replacement API is moved to a different object, developers create that object before calling the replacement API.
    
    \begin{lstlisting}[caption={Usage update pattern: Same return type and parameters.}, label=lstcat1rule]
if (Build.VERSION.SDK_INT >= specific_version) {
    replacementAPI();
} else {
    deprecatedAPI();
}
    \end{lstlisting}
    
    If the replacement API requires additional initialization (e.g., a context or configuration object not needed by the deprecated API), the simple version check may fail without proper setup, complicating automation. Listing \ref{lstcat1edge} shows a solution with initialization.
    
    \begin{lstlisting}[caption={Usage update pattern: Same return type and parameters with new object initialization for replacement API.}, label=lstcat1edge]
if (Build.VERSION.SDK_INT >= specific_version) {
    ReplacementObject obj = new 
                ReplacementObject(context);
    obj.replacementAPI();
} else {
    deprecatedAPI();
}
    \end{lstlisting}

    \item \textbf{Same return type with different parameters.} 
    In this scenario, developers typically adapt the method call based on the Android version, introducing or rearranging parameters for the replacement API. 
    
    \begin{lstlisting}[caption={Usage update pattern: Same return type with different parameters.}, label=lstcat2rule]
if (Build.VERSION.SDK_INT >= specific_version) {
    replacementAPI(param1, param2, additionalParam);
} else {
    deprecatedAPI(param1, param2);
}
    \end{lstlisting}
    
    Listing \ref{lstcat2rule} shows the usage update pattern for this category. Automatically determining values for new parameters, especially when they lack direct correspondence to old ones, can be challenging. Parameter order changes also demand programmatic discernment.

    \item \textbf{Different return type and parameters.} 
    Developers often encapsulate each API in separate methods for this category, handling return type conversions to maintain consistent behavior. 
    
\begin{lstlisting}[caption={Usage update pattern: Different return type and parameters.}, label=lstcat3rule]
ReturnType result;
if (Build.VERSION.SDK_INT >= specific_version) {
    NewType newResult = replacementAPI(param1);
    result = convertType(newResult);
} else {
    result = deprecatedAPI(param2);
}
\end{lstlisting}
    
    Listing \ref{lstcat3rule} shows the usage update pattern for this category. Many deprecated APIs here map to multiple replacements (e.g., \url{ConnectivityManager.getAllNetworkInfo()} to \url{getAllNetworks()} and \url{getNetworkInfo(android.net.Network)}), with \url{convertType()} often guided by deprecation messages. Automation requires handling return type conversions and semantic shifts, such as deciding between multiple replacements, which demands deep contextual understanding.

    \item \textbf{API fields of the same data type.} 
    For simple field name changes, developers use version-based conditionals to fetch the correct field, ensuring smooth transitions.
    
\begin{lstlisting}[caption={Usage update pattern: API fields of the same data type.}, label=lstcat4rule]
Type value = 
    (Build.VERSION.SDK_INT >= specific_version) ? 
             ReplacementField : DeprecatedField;
\end{lstlisting}
    
    Listing \ref{lstcat4rule} shows the usage update pattern for this category. Fields may carry different semantics or value ranges in the new version, challenging automation.
    
    If the replacement field is part of a different object, developers create that object before accessing the field, similar to Listing \ref{lstcat1edge} of Category 1.

    \item \textbf{API fields of different datatypes.} 
    Developers commonly use conversions or utility methods when data types change between versions to keep apps robust.
    
\begin{lstlisting}[caption={Usage update pattern: API fields of different datatypes.}, label=lstcat5rule]
Type value;
if (Build.VERSION.SDK_INT >= specific_version) {
    NewType newValue = ReplacementField;
    value = convertType(newValue);
} else {
    value = DeprecatedField;
}
\end{lstlisting}
    
    Listing \ref{lstcat5rule} shows the update pattern for this category. Automation faces complexity beyond type conversion, needing to understand field relationships to avoid behavioral disruptions.

    \item \textbf{Different types.} 
    Generalizing this category is difficult, as transitions between fields and methods require careful API access handling and possible conversion. Listing \ref{lstcat6rule} shows the common field-to-method pattern; if the replacement returns the same data type, \texttt{convertType()} may be unnecessary.
    
\begin{lstlisting}[caption={Usage update pattern: Different types.}, label=lstcat6rule]
Type value;
if (Build.VERSION.SDK_INT >= specific_version) {
    NewType newValue = ReplacementAPI();
    value = convertType(newValue);
} else {
    value = DeprecatedField;
}
\end{lstlisting}
    
    Automation must grasp usage context to preserve intent, as methods may introduce side effects absent in fields, or vice versa.

    If the replacement method requires asynchronous execution (e.g., a callback), the synchronous field access pattern fails. Listing \ref{lstcat6edge} adapts to asynchronous calls.
    
\begin{lstlisting}[caption={sage update pattern: Different types with asynchronous replacement API.}, label=lstcat6edge]
Type value;
if (Build.VERSION.SDK_INT >= specific_version) {
    ReplacementAPI(new Callback() {
        @Override
        void onResult(NewType newValue) {
            value = convertType(newValue);
        }
    });
} else {
    value = DeprecatedField;
}
\end{lstlisting}
    
\end{enumerate}

\section{Applicability Study}
\label{sec:applicability}

In this study, we aim to answer the following research question:

\textit{To what extent are the identified API usage-update patterns useful for improving LLM-based deprecated API migration?}

We evaluate the applicability of the patterns using the GUPPY benchmark~\cite{mahmud2025automated}, which contains 156 deprecated API update tasks collected from Android API levels 33 and 34. GUPPY provides the deprecated API usage, the API level in which it was deprecated, the recommended replacement API, and the enclosing source-code context for each task. We use only the initial zero-shot update generated by GUPPY and do not use its test generation or iterative refinement components.

We compare the original GUPPY zero-shot prompt with a pattern-guided version. The original prompt provides the deprecated API, deprecation level, replacement API, and source-code context, and asks the model to maintain compatibility across Android versions. We extend this prompt by adding the migration category and the corresponding usage-update pattern identified in our empirical study. The updated prompt is:

\begin{tcolorbox}[
    colback=gray!10,
    colframe=gray!50,
    boxrule=0.5pt,
    arc=2pt,
    left=6pt,
    right=6pt,
    top=6pt,
    bottom=6pt
]
\small
You are tasked with updating a deprecated Android API usage.
Follow these steps:

1. Identify the deprecated API method \texttt{[Deprecated API]} in the code.

2. The API was deprecated in API level \texttt{[API level]}.

3. Review the recommended replacement API(s):
\texttt{[Replacement API]}.

4. This migration belongs to \texttt{[Migration Category]}.

5. Use the following API usage-update pattern as guidance:

\texttt{[Usage-Update Pattern]}

6. Adapt the pattern to the provided source-code context. Update the code to use
the replacement API while maintaining backward compatibility across Android
versions, using version checks when necessary.

7. Preserve the original behavior of the code and provide only the updated code
segment.

\texttt{[Code snippet]}
\end{tcolorbox}

For each of the 156 tasks, we generate one update using the original GUPPY prompt and one using the pattern-guided prompt. We manually assess whether each generated update correctly uses the recommended replacement API, maintains compatibility across relevant Android versions, and preserves the intended behavior of the original code. We then compare the overall and category-level numbers of correct updates produced by the two prompts.

Table~\ref{tab:applicability-results} presents the results of applying the
identified usage-update patterns to the 156 API update tasks in the GUPPY
benchmark~\cite{mahmud2025automated}. The original GUPPY zero-shot prompt
correctly updated 139 of the 156 API usages. The pattern-guided prompt correctly
updated 147 usages.

\begin{table}[t]
\centering
\caption{Comparison of the original and pattern-guided GUPPY prompts.}
\label{tab:applicability-results}
\begin{tabular}{lrrr}
\hline
\textbf{Category} &
\textbf{\# APIs} &
\textbf{GUPPY} &
\textbf{With Pattern} \\
\hline
Category 1 & 17 & 16 & 17 \\
Category 2 & 55 & 50 & 54 \\
Category 3 & 40 & 35 & 37 \\
Category 4 & 8  & 7  & 8  \\
Category 5 & 13 & 12 & 13 \\
Category 6 & 23 & 19 & 18 \\
\hline
\textbf{Total} & \textbf{156} & \textbf{139} & \textbf{147} \\
\hline
\end{tabular}
\end{table}

The pattern-guided prompt improves the number of correct updates for Categories
1--5. Overall, it produces 147 correct updates, compared with 139 correct
updates from the original GUPPY zero-shot prompt. This result suggests that
the usage-update patterns derived from real-world developer changes provide
useful guidance for LLM-based API migration.

Category 6 is the only category for which the pattern-guided prompt performs
slightly worse, producing 18 correct updates compared with 19 for the original
prompt. To investigate this result, we revisited the real-world Category 6
updates used to derive the pattern. We found that the selected usage-update
pattern occurs in only 76\% of these updates. The remaining 24\% follow other
update strategies that are not represented by the supplied pattern. On the other hand, patterns in Categories
1--5 occurs in more than 93\% of the updates we analyzed.  Therefore,
a single generalized pattern is insufficient to represent the diverse migration
strategies within Category 6.
\section{Implications} \label{discussion}

Our findings provide practical guidance for researchers developing automated API migration techniques and for Android developers maintaining applications across API levels.

\subsection{For Researchers}

\subsubsection*{\textbf{Automated Deprecated API Usage Update}}

The identified migration categories and usage-update patterns provide reusable guidance for automated migration approaches. Simpler categories, such as migrations with the same return type and parameters, can often be handled through direct replacement, version checks, and limited object initialization. More complex categories require parameter construction, return-type conversion, multiple replacement APIs, or changes in interaction mechanisms.

Our applicability study further shows that these patterns can improve LLM-based API usage update. The pattern-guided prompt correctly updated 147 of the 156 GUPPY tasks, compared with 139 using the original GUPPY~\cite{mahmud2025automated}. This result suggests that empirically derived migration knowledge can provide useful domain guidance to LLMs. However, the Category 6 pattern appeared in only 76\% of the corresponding real-world updates, indicating that heterogeneous categories may require multiple specialized patterns or additional program context.

\subsubsection*{\textbf{API Evolution Studies}}

The proposed categories provide a structured framework for studying how deprecated APIs differ from their replacements. They can support future analyses of migration complexity, recurring API design changes, and the types of transformations that are most suitable for automation.

\subsection{For Android App Developers}

The migration categories and usage-update patterns can help developers understand the changes required when adopting replacement APIs, including version checks, parameter adaptation, object initialization, and data conversion. They can also help identify complex migrations that require additional manual review.

API providers and open-source communities can further improve migration support by providing clear replacement information and representative update examples, particularly for migrations involving multiple APIs, callbacks, or changes in API interaction mechanisms.
\section{Threats to Validity} \label{threats}

Regarding external validity, our study utilized 1,000 randomly selected Android APIs and 4,336 update examples of these APIs from GitHub. To mitigate this threat, we employed a random sampling strategy and assessed the sample’s representativeness by comparing category distributions against trends in the full list of Android APIs when checking the distribution of the categories.

{The primary threats to internal validity pertain to potential errors in our algorithm's implementation for collecting data from the Android developers' website. To mitigate this threat, we tested our implementations on examples that we could manually verify.
An additional threat to internal validity arises in RQ2, where manual analysis of 4,336 update examples could introduce subjectivity in rule derivation. We mitigated this by involving two experienced authors and using a structured, consensus-driven process, though unexamined developer intent or rare edge cases might limit rule generalizability. To enhance robustness, we quantified rule coverage, but the sample size and GitHub bias may not fully capture all real-world practices. Future work with larger, diverse datasets could further validate these findings.}

Where threats to construct validity are concerned, we have collected the deprecated APIs from the Android developers website, which misses some hidden deprecated APIs. To mitigate this threat, we have collected the deprecated APIs available in the latest version of the Android API level.
\section{Related Work} \label{relatedWork}
In this section, we discuss the related works on API deprecation and evolution studies.

\subsection{API Deprecation}
As software systems evolve, APIs are often deprecated to introduce new functionality or to address security or performance concerns. Ideally, deprecated APIs should follow a deprecate-replace-remove cycle, where the API is first marked as deprecated, replaced by a new API, and eventually removed from the source code base \cite{10.1145/2950290.2950298,1510134,10.5555/1133105.1133107,10.1145/1932682.1869518}. 
However, many deprecated APIs remain in use for years despite being marked as such. Zhou et al. \cite{10.1145/2950290.2950298} found that in open-source Java frameworks and libraries, the traditional deprecate-replace-remove cycle is often not followed, leading to outdated code examples on the web and confusion for developers.

Sawant et al. \cite{8529833} found that the current Java deprecation mechanism does not meet developers' needs. Raemaekers et al. \cite{6975655} found that deprecated APIs are sometimes not removed from Java systems, leading to inconsistencies in code bases like the Android framework. Brito et al. \cite{7476657} and Ko et al. \cite{7091210} emphasize the significance of clear replacement messages in facilitating the adoption of new APIs.
Xi et al. \cite{10.1145/3361242.3361246} developed Dep2Doc, a tool to help developers migrate from deprecated to documented APIs. Li et al. \cite{li2018characterising} analyzed Android's codebase and over 13,000 apps to understand deprecated API usage. While this study offers insights into deprecated API impacts, our work concentrates on how the deprecated API usages are updated by examining API migration and common usage update rules.

There are several works on the automatic update of deprecated API usage in Android \cite{lamothe2020a3,fazzini2019automated,haryono2020automatic,haryono2021androevolve,zhao2022towards,thung2019towards,thung2020automated}. 
All these methods require developers to know the usage of deprecated APIs and replacement APIs, which can be eased by our study in this paper. 
Haryono et al. \cite{9609153} developed MLCatchUp, which utilizes natural language processing to suggest replacements for deprecated machine learning APIs in Python based on semantic similarity. Our study focuses on the API migration of deprecated and replacement Android APIs and their usage update patterns.

\subsection{API Evolution}
Previous studies have explored the development of Android APIs, examining various aspects of API evolution and its impact. For example, Lamothe et al. \cite{lamothe2021systematic} conducted a systematic review of API evolution research, identifying six primary research areas that focus on API evolution. Linares et al. \cite{linares2013api,linares2014api} found that successful apps tend to use APIs that are less prone to faults and changes and Android API changes lead to a higher number of Stack Overflow discussions.

McDonnell et al. \cite{mcdonnell2013empirical} analyzed Android APIs using version history data from GitHub and found that API updates are more vulnerable than other changes. Similarly, Mutchler et al. \cite{mutchler2016target} evaluated older Android apps on newer devices for security concerns, while Li et al. \cite{li2016accessing} tracked inaccessible Android APIs by examining the code base. 
Yang et al. \cite{Yang:2018:AOS:3197231.3197258} analyzed the impact of Android updates on apps, and Mahmud et al. \cite{apicia-jss} assessed the effects of changing the target API on app source code and tests. 
Researchers have also investigated the evolution of APIs on other platforms. For instance, Businge et al. \cite{businge2012survival,businge2013analyzing,businge2015eclipse,6079836} conducted multiple empirical investigations on the Eclipse platform and AWT/Swing API evolution.

Several techniques \cite{ficfinder,li2018cid,huang2018understanding,he2018understanding,pivot,acid,mahmud2022field} have been proposed to detect API evolution-induced compatibility issues. 
Unlike these techniques, our study analyzes the deprecated APIs that cause compatibility issues and their migration and usage updates.
\section{Conclusion} \label{conclusion}

As the Android ecosystem evolves, API deprecation continues to create compatibility and maintenance challenges for developers. This study systematically examines deprecated Android APIs, their recommended replacements, and the ways developers update their usages in practice. We identify six migration categories and derive recurring usage-update patterns involving version checks, parameter adaptation, object initialization, data-type conversion, and changes in API interaction mechanisms.

We further evaluate the practical usefulness of these patterns using the GUPPY benchmark~\cite{mahmud2025automated}. The pattern-guided prompt improves the number of correct zero-shot updates from 139 to 147 out of 156 tasks, demonstrating that empirically derived migration knowledge can strengthen LLM-based API updating. At the same time, our analysis shows that a single pattern may not adequately represent heterogeneous migration categories, highlighting the need for multiple specialized patterns and richer program context for complex cases. Overall, the proposed categories and patterns provide a practical foundation for improving automated API migration and supporting developers in maintaining reliable Android applications.


\balance
\bibliographystyle{plain}
\bibliography{Bibliography}

\end{document}